# Phase controlled light switching at low power levels

Hoonsoo Kang, Gessler Hernandez, Jiepeng Zhang, and Yifu Zhu

Department of Physics

Florida International University

Miami, Florida  33199

Abstract

We report experimental observations of interference between three-photon and one-photon excitations, and phase control of light attenuation/transmission in a four-level system. Either constructive interference or destructive interference can be obtained by varying the phase and/or frequency of a weak control laser. The interference enables absorptive switching of one field by another field at different frequencies and ultra-low light levels.

Laser induced interference plays an important role in interactions of radiation and matter, and has found numerous applications in optical physics. One such example is electromagnetically induced transparency (EIT) [1]. EIT has been used to obtain slow light speed in an absorbing medium [2] and to study nonlinear optics at low light levels down to single photons [3]. It has been shown that the EIT technique may be used to explore quantum information science [4].

Light switching light at low light levels based on quantum interference or other mechanisms have been studied in recent years and quantum light switching with single photons may have important applications in quantum electronics [5]. Here we present a scheme based on phase-controlled quantum interference that may be used to realize a single-photon light switch. We report experimental observations of interference between three-photon and one-photon excitations in a four-level system and the resulting phase switching of light absorption/ transmission at low light powers ($<10^{-7}$ W). We show that the interference leads to interesting spectral and dynamic features, and one weak field can be used to control another weak field and vice versa at ultra-low light levels.

Before proceeding further, we note that Georgiades et al observed quantum interference of two photon transitions in cold atoms [6]; Korsunsky et al studied the phase dependent coherent population trapping (CPT) [7]; Huss et al reported correlation of the phase fluctuation in a double-$\Lambda$ system [8]; Deng and Payne showed that EIT can be induced by destructive interference between three-photon and one-photon excitation channels and a pair of matched light pulses can be generated in a three-level medium [9]. Also a variety of other phenomena and applications involving three or four-level EIT systems have been studied in recent years [10-21].

Consider a four-level system driven by two coupling fields (Fig. 1). The coupling field 1 (2) drives the transition |2>-|3> (|4>) with Rabi frequency $\Omega_1$ ($\Omega_2$). Two weak fields, one as probe and another as control, drive the transitions |1>-|3> and |1>-|4> with Rabi frequency $\Omega_p$ and $\Omega_c$ (the probe and control are

interchangeable). Here $\Omega_i = |\Omega_i| e^{i\phi_i}$ (i=1,2, p, and c) is characterized by the amplitude $|\Omega_i|$ and phase $\phi_i$.

The frequency detunings for the respective transitions are defined as $\Delta_p = \omega_p - \omega_{31}$, $\Delta_1 = \omega_1 - \omega_{32}$, $\Delta_c = \omega_c - \omega_{41}$, and $\Delta_2 = \omega_2 - \omega_{42}$ ($\omega_i$ (i=p, 1, c, 2) is the angular frequency of the laser field i). For $|\Omega_1| \sim |\Omega_2|$, $|\Omega_p| \sim |\Omega_c|$, $|\Omega_{1(2)}| >> |\Omega_{c(p)}|$, and $\Delta_i = 0$ (i=p,1,c, and 2), the adiabatic excited-state populations are

$$P_3 = \left| \frac{\Omega_p(|\Omega_2|^2 + \gamma_2\gamma_4) - \Omega_1\Omega_2^*\Omega_c}{\gamma_2\gamma_3\gamma_4 + \gamma_3|\Omega_2|^2 + \gamma_4|\Omega_1|^2} \right|^2, \quad (1a) \quad \text{and} \quad P_4 = \left| \frac{\Omega_c(|\Omega_1|^2 + \gamma_2\gamma_3) - \Omega_2\Omega_1^*\Omega_p}{\gamma_2\gamma_3\gamma_4 + \gamma_3|\Omega_2|^2 + \gamma_4|\Omega_1|^2} \right|^2, \quad (1b)$$

where $\gamma_2$ is the decay rate of the ground-state coherence $\rho_{12}$. The first term in Eq. (1a) ((1b)) represents the one-photon excitation $|1\rangle$-$|3\rangle$ ($|1\rangle$-$|4\rangle$) while the second term represents the three-photon excitation $|1\rangle$-$|4\rangle$-$|2\rangle$-$|3\rangle$ ($|1\rangle$-$|3\rangle$-$|2\rangle$-$|4\rangle$). The two excitation paths interfere with each other, which can be manipulated by varying the phases and amplitudes of the laser fields. When $\Omega_1\Omega_c = \Omega_2\Omega_p$ (neglecting $\gamma_2$, which is justified due to $|\Omega_{1(2)}| >> \gamma_2$ and $\gamma_{3(4)} >> \gamma_2$), the interference is destructive: $P_3$ and $P_4$ vanish, and the probe and control fields propagate in the medium without attenuation. When $\Omega_1\Omega_c = -\Omega_2\Omega_p$, the interference is constructive: $P_3$ and $P_4$ are maximized, and the probe and control fields are attenuated in the medium. The four-level system can be used for absorptive switching of one weak field by another weak field. To discuss the interference switching, we consider propagation of the probe and control fields in the four-level medium of length $\ell$. For $\Omega_1 \sim \Omega_2$, $\Omega_p \sim \Omega_c$, $\Omega_{1(2)} >> \Omega_{c(p)}$, $\Delta_c = \Delta_p$, and the four laser fields propagate in the z direction (neglecting depletion of the two coupling fields), the Maxwell equations for the two weak fields are

$$\frac{d\Omega_p}{dz} = \frac{iK_{13}(|\Omega_1|^2 - \delta_1\delta_c)}{\Lambda}\Omega_p - \frac{iK_{13}\Omega_1\Omega_2^*}{\Lambda}\Omega_c \quad (2a)$$

$$\frac{d\Omega_c}{dz} = \frac{iK_{14}(|\Omega_2|^2 - \delta_1\delta_p)}{\Lambda}\Omega_c - \frac{iK_{14}\Omega_1^*\Omega_2}{\Lambda}\Omega_p. \quad (2b)$$

Here $\delta_p = \Delta_p + i\gamma_3$, $\quad \delta_1 = \Delta_p - \Delta_1 + i\gamma_2$, $\quad \delta_c = \Delta_c + i\gamma_4$, and $\quad \Lambda = \delta_1\delta_2\delta_p - \delta_p|\Omega_2|^2 - \delta_2|\Omega_1|^2$, and

$K_{ij} = \dfrac{2\pi N \omega_{ij}|\mu_{ij}|^2}{\hbar c}$ (N is the atomic density).

Eq. (2a) and (2b) can be solved analytically. Fig. 2 plots the calculated probe transmission versus $\Delta_p$ in the four-level system. Without the control field (Fig. 2(a)), the probe field is attenuated in the medium with peak absorptions at $\Delta_p \sim \pm\Omega = \pm\sqrt{\Omega_1^2 + \Omega_2^2}$ and $\Delta_p=0$ (EIT enhanced nonlinear absorption [5]). When the control field is present and $\Omega_p(0)\Omega_2 = \Omega_c(0)\Omega_1$ ($\Omega_{p(c)}(0)$ is the incident probe (control) Rabi frequency at z=0), the absorption for the probe and the control at $\Delta_p=\Delta_c=0$ are suppressed (destructive interference) while the absorption for the probe and the control at $\Delta_p=\Delta_c\approx\pm\Omega$ are enhanced (constructive interference) (Fig. 2b). That is, EIT is induced simultaneously for both the probe and control fields at $\Delta_p=\Delta_c=0$. The group velocity of the two weak fields are given by $V_{g(p)} = c/(1 + \dfrac{\omega_p}{2}\dfrac{\gamma_4 K_{13}}{\gamma_3|\Omega_2|^2 + \gamma_4|\Omega_1|^2})$, and $V_{g(c)} = c/(1 + \dfrac{\omega_c}{2}\dfrac{\gamma_3 K_{14}}{\gamma_3|\Omega_2|^2 + \gamma_4|\Omega_1|^2})$. Matching of the slow group velocities, $V_{g(p)} \approx V_{g(c)}$, is possible for atomic systems in which $\omega_p\gamma_4 K_{13} \approx \omega_c\gamma_3 K_{14}$ (such as the Rb system of Fig. 3), and the four-level system may be used to produce a pair of matched light pulses [9]. On the other hand, when the four laser fields satisfy the condition $\Omega_p(0)\Omega_2 = -\Omega_c(0)\Omega_1$, the probe and control absorptions at $\Delta_p=\Delta_c=0$ are enhanced (constructive interference) while the probe and control absorptions at $\Delta_p=\Delta_c\approx\pm\Omega$ are suppressed (destructive interference) (Fig. 2(c)). We note that the calculations are valid for arbitrarily weak probe and control fields. The interference requires the control Rabi frequency satisfying $|\Omega_c(0)|=|\Omega_p(0)\Omega_2/\Omega_1|$. With $|\Omega_c(0)|=\sqrt{n_c}g_c$ and $|\Omega_p(0)|=\sqrt{n_p}g_c$ ($n_{c(p)}$ is the average number (expectation value) of the control (probe) photons and $g_{c(p)}$ is the coupling coefficient), the amplitude condition becomes $n_c = |\Omega_2 g_p/(\Omega_1 g_c)|^2 n_p = \alpha n_p$ ($\alpha$ can be

≥1 or <1). Thus the phase controlled interference may be implemented near single photon levels ($n_c \sim n_p \sim 1$) and the four-level system may be used as an absorptive quantum switch, which turns on or off of single probe photons by single control photons at different frequencies. We note that when the photon number approaches one, the quantum fluctuation is important and understanding of the statistical properties of the photon switching requires calculations of the second-order noise correlations in the four-level system.

The dressed state picture provides a simple explanation for the interference of the two weak fields in the four-level system. The two resonant coupling fields create a manifold of three dressed states, the semi-classical representation of which is given by $|+> = \frac{1}{\sqrt{2}}\{|1> - \frac{\Omega_1}{\Omega}|3> - \frac{\Omega_2}{\Omega}|4>\}$, $|0> = \frac{\Omega_2}{\Omega}|3> - \frac{\Omega_1}{\Omega}|4>$, and $|-> = \frac{1}{\sqrt{2}}\{|1> + \frac{\Omega_1}{\Omega}|3> + \frac{\Omega_2}{\Omega}|4>\}$ ($\Omega = \sqrt{\Omega_1^2 + \Omega_2^2}$). The corresponding level shifts are $\Omega$, 0, and $-\Omega$, respectively. With two weak fields $\Omega_c$ and $\Omega_p$, the transition probability from the state $|1>$ to the dressed states $|+>$ and $|->$ is $P_{1\pm} \propto |\Omega_p\Omega_1 + \Omega_c\Omega_2|^2 = |\Omega_p|^2|\Omega_1|^2 + |\Omega_c|^2|\Omega_2|^2 + 2|\Omega_p||\Omega_1||\Omega_c||\Omega_2|\cos(\Phi_2 + \Phi_c - \Phi_1 - \Phi_p)$, and the transition probability from the state $|1>$ to the dressed state $|0>$ is $P_{10} \propto |\Omega_p\Omega_2 - \Omega_c\Omega_1|^2 = |\Omega_p|^2|\Omega_2|^2 + |\Omega_c|^2|\Omega_1|^2 - 2|\Omega_p||\Omega_1||\Omega_c||\Omega_2|\cos(\Phi_2 + \Phi_c - \Phi_1 - \Phi_p)$. $\Phi_i$ (i=1,2,c, and p) is the phase of the laser field. The interference between the two excitation paths can be manipulated by varying the amplitudes and phases of the four laser fields. For example, when $\Omega_1 = \Omega_2$ and $\Omega_c = \Omega_p$, complete destructive interference occurs for the transitions $|1>- |0>$ while complete constructive interference occurs for the transitions $|1>-|\pm>$.

Our experiment is done with cold $^{85}$Rb atoms confined in a magneto-optical trap (MOT) described in our earlier studies [11]. A simplified experimental set up with two frequency-modulated lasers for the spectral measurements is depicted in Fig. 3(a). An extended-cavity diode laser with a beam diameter ~ 3 mm and output power ~ 50 mW is used as the coupling laser. The driving electric current to the diode laser is modulated at $\delta$=181 MHz with a modulation index ~0.5, which produces two first-order frequency sidebands separated by 362 MHz. The two sidebands are tuned to the $^{85}$Rb $D_1$ F=3→F'=2 and F=3→F'=3 transitions respectively and serve as the two coupling fields ($\Omega_1 \approx -\Omega_2$ due to a π phase difference between the two sidebands). Another extended-cavity diode laser with a beam diameter ~ 0.4 mm and output power attenuated to ~ 0.1 mW is also current modulated at 181 MHz and the two first-order sidebands are tuned to the $D_1$ F=2→F'=2 and F=2→F'=3 transitions, serving as the probe field and control field respectively ($\Omega_c \approx -\Omega_p$). The carrier and the higher-order sidebands of the coupling laser and the weak laser are detuned from the atomic transitions by at least 181 MHz and their effects on the four-level system can be neglected at the laser intensity levels used in the experiment. The coupling laser and the weak laser are circularly polarized ($\sigma^+$) and interact with the Rb transitions to form 4 separate sets of the double-Λ type four-level system among the magnetic sublevels. The two lasers propagate in the same direction separated by a small angle (~1°) and overlapped inside the MOT. The transmitted beam of the weak laser (both the probe and the control) is collected by a photodetector and the fluorescence photons from spontaneous emissions of the atoms in the excited states |3> and |4> are collected by another photodetector after the imaging optics.

The experiment is run in a sequential mode with a repetition rate of 5 Hz. all lasers are turned on or off by Acousto-Optic Modulators (AOM) according to the time sequence described below. For each period of 200 ms, ~198 ms is used for cooling and trapping of the $^{85}$Rb atoms, during which the trapping laser and the repump laser are turned on by two AOMs while the coupling laser and the weak laser are off. The time for the data collection lasts ~ 2 ms, during which the trapping laser and the repump laser are turned off as well as the current to the anti-Helmholtz coils of the MOT, and the coupling laser and the weak laser are turned on. For the spectral measurements, the weak laser frequency is scanned across the $^{85}$Rb $D_1$ F=2→F' transitions after a 0.1 ms delay and the transmission of the weak laser and the fluorescent photons (proportional to the excited-state population) are then recorded.

The light transmission and the fluorescence intensity versus the weak-laser detuning ($\Delta_c = \Delta_p$) are plotted in

Fig. 4. Fig. 4 (a) and (b) show the measurements without the control field (the weak laser is not frequency modulated) and represents the EIT manifested absorption spectra in the four-level system observed before [5]. The light transmission spectrum with both the probe and control fields present (the weak laser is frequency modulated) is plotted in Fig. 4(c), and shows that the absorption at the resonance ($\Delta_p=\Delta_c=0$) is suppressed by the destructive interference, which results in simultaneous EIT for both the probe field and the control field. Correspondingly, the measured fluorescence intensity is suppressed at $\Delta_p=\Delta_c=0$ (Fig. 4(d)), which demonstrates the suppression of the excited-state population by the destructive interference between three-photon and one-photon excitations.

We studied the phase switching of the probe absorption/transmission in the four-level system with an experimental set up shown in Fig. 3(b). With an AOM, we obtain two first-order beams of the weak laser: one as the probe (the frequency shifted down by $\delta$) and another as the control (the frequency shifted up by $\delta$). The up-shifted control beam is then passed through an electro-optic modulator (EOM, New Focus 4002) and its phase is varied by a voltage applied to the EOM. The two beams with about equal powers are combined in a beam combiner and then are coupled into a polarization-maintaining single-mode fiber, the output of which are collimated, attenuated to a power level of $< 10^{-7}$ W (the intensity~0.1 mW/cm$^2$), and directed to be overlapped with the frequency-modulated coupling laser in the MOT. The transmission of the combined probe and control beams is collected by a photodetector. The left panel of Fig. 5 ((a) and (b)) plots the transmitted probe and control beams versus the control laser phase $\Phi_c$ as $\Phi_c$ is varied by a sinusoidal voltage applied to the EOM (Fig. 5c). Fig. 5(a) plots the light transmission versus $\Phi_c$ at $\Delta_p=\Delta_c\approx 0$ and Fig. 5(b) plots the light transmission versus $\Phi_c$ at $\Delta_p=\Delta_c\approx\Omega$. The data show that there is a $\pi$ phase difference in the interference pattern between the two cases, illustrating the phase and frequency control of the light transmission by the quantum interference. The right panel of Fig. 5 ((a') and (b')) plots the transmitted light intensity versus time when a square-wave voltage is applied to the EOM to switch $\Phi_c$ from 0 to $\pi$. The data

4show the periodic switching of the light transmission versus $\Phi_c$ and the switching pattern is reversed through the $\pi$ phase reversal at different frequencies. The switching efficiency may be defined as $\eta = (I_{close} - I_{open})/I_{in}$. Here $I_{in}$ is the incident light intensity, $I_{close}$ is the transmitted intensity when the switch is closed, and $I_{open}$ is the transmitted intensity when the switch is open. For perfect switching, $\eta$=100% ($I_{close}$= $I_{in}$ and $I_{open}$=0). In our experiments, when the switch is open, the light transmission is ≈20%, which is limited by the optical depth of the cold atomic cloud; when the switch is closed, the light transmission is ≈80%, which is limited by the absorption loss due to the laser frequency drifts, the Zeeman broadening from the residual magnetic field, and the decay rate $\gamma_2$ of the ground state coherence. The observed switching efficiency is $\eta$≈60% for $\Delta_p$=$\Delta_c$≈0 and $\eta$≈55% for $\Delta_p$=$\Delta_c$≈$\Omega$.

In conclusion, we have observed interference of three-photon and one-photon excitations in a four-level system, and demonstrated the phase and frequency control of the absorptive switching at low light levels (~0.1 mW/cm$^2$). The four-level phase control scheme can be used to produce matched slow light pulses and may lead to realistic implementation of a quantum switch in which the attenuation/transmission of single photons is controlled by single photons at different frequencies.

4This work is supported by the National Science Foundation.

Figure Captions

Fig. 1 (a) Four-level system. $\gamma_3$ ($\gamma_4$) is the spontaneous decay rate ($\gamma_3 \approx \gamma_4 = 2\pi \times 5.4 \times 10^6$ s$^{-1}$). Interference occurs between (b) three-photon excitation |1>-|3>-|2>-|4> and (c) one-photon excitation |1>-|4>. The interference between three-photon excitation |1>-|4>-|2>-|3> and one-photon excitation |1>-|3> is not drawn here.

Fig. 2. (a) Calculated probe transmission versus $\Delta_p$ without the control laser. (b) Calculated probe transmission versus $\Delta_p = \Delta_c$ with $\Omega_c(0) = \Omega_p(0)$. The absorption is suppressed by destructive interference at $\Delta_p = 0$ and enhanced by constructive interference at $\Delta_p = \pm \Omega$. (c) Calculated probe absorption versus $\Delta_p = \Delta_c$ with $\Omega_c(0) = -\Omega_p(0)$. The absorption is enhanced by constructive interference at $\Delta_p = 0$ and suppressed by destructive interference at $\Delta_p = \pm \Omega$. The parameters are $\Omega_1 = \Omega_2 = \gamma_3$, $\gamma_3 = \gamma_4$, $\gamma_2 = 0.02\gamma_3$, $\Delta_1 = \Delta_2 = 0$, and $K_{13}\ell/\gamma_3 = K_{14}\ell/\gamma_4 = 1$.

Fig. 3. (a) Simplified diagram of the experimental set up with two frequency-modulated lasers and the coupled four-level $^{85}$Rb system. (b) Simplified diagram of the experimental set up used for measurements of the light transmission versus the phase variation of the control field. AOM: acousto-optic modulator; EOM: electro-optic modulator; PMF: polarization maintaining fiber; λ/4: quarter-wave plate; DL: extended-cavity diode laser; M: mirror; D: photodetector.

Fig. 4 Measured probe transmission ((a) and (c)) and measured fluorescence intensity ((b) and (d)) versus the detuning $\Delta_c = \Delta_p$. Solid (dashed) lines are experimental data (calculations). In (a) and (b), the weak laser is not frequency modulated (no control field component). In (c) and (d), the weak laser is frequency modulated to produce two sidebands used as the probe and control fields, and the absorption peaks at $\Delta_p = 0$ in (a) and (b) are suppressed by the destructive interference. The experimental parameters are $\Omega_1/(2\pi) \approx \Omega_2/(2\pi) \approx 4$ MHz, $\Omega_p/(2\pi) \approx \Omega_c/(2\pi) \approx 0.2$ MHz and $\Delta_1 = \Delta_2 = 0$. The fitting parameters are $\gamma_2 \approx 0.02\gamma_3$,

$K_{13}\ell/\gamma_3 = K_{14}\ell/\gamma_4 = 0.8$.

Fig. 5. Transmission of the control and the probe fields versus the phase variation of the control field. Dots are experimental data and solid lines are calculations with the adiabatic phase change. Left panel: $\Phi_c = \pi\cos(2\pi ft) + \varphi_0$ (f=2.3 KHz and $\varphi_0$ is set by the DC off-set voltage); right panel: square-wave shift of $\Phi_c$ (between 0 and $\pi$) at f=2.3 KHz. (a) and (a'): the light transmission at $\Delta_c=\Delta_p\approx 0$; (b) and (b'): the light transmission at $\Delta_c=\Delta_p\approx\Omega$; (c) and (c'): the control-laser phase $\phi_c$ versus time. $\Omega_1/(2\pi)\approx\Omega_2/(2\pi)\approx 4.5$ MHz, and the other parameters are the same as that in Fig. 4.

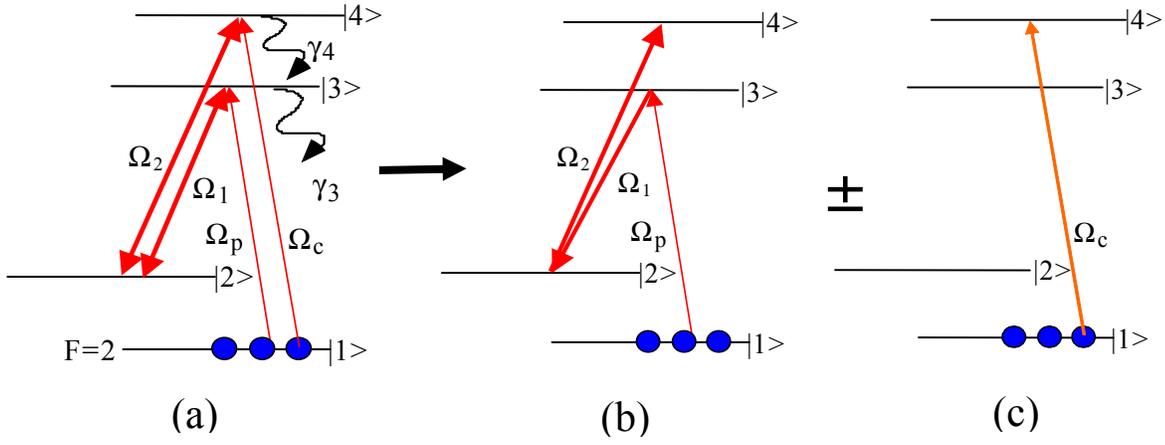

Fig. 1 Kang et al

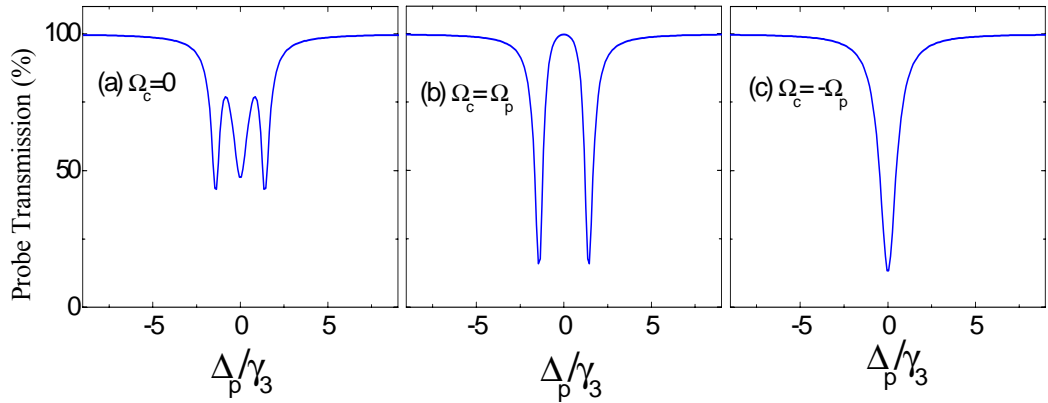

Fig. 2 Kang et al

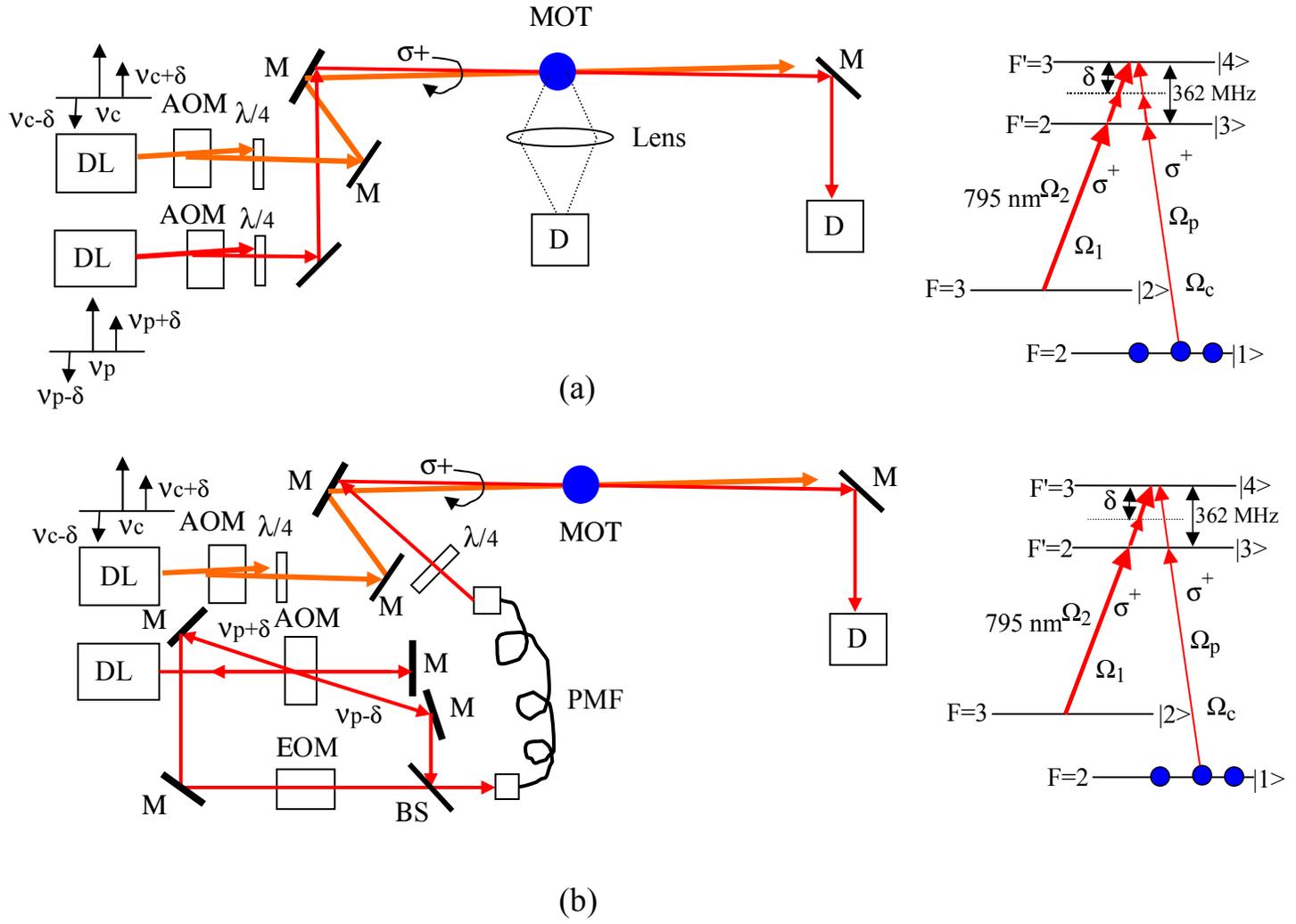

Fig. 3 Kang et al

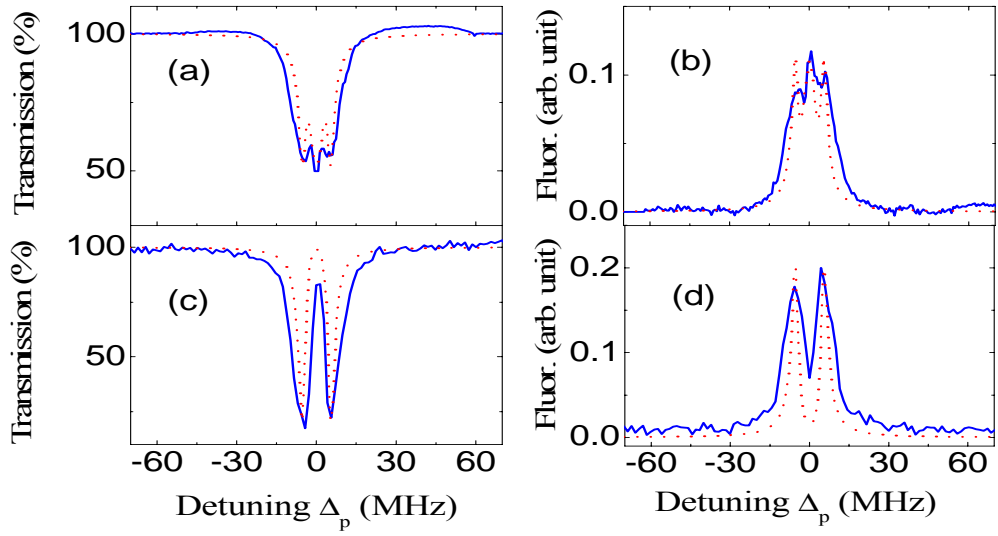

Fig. 4   Kang et al

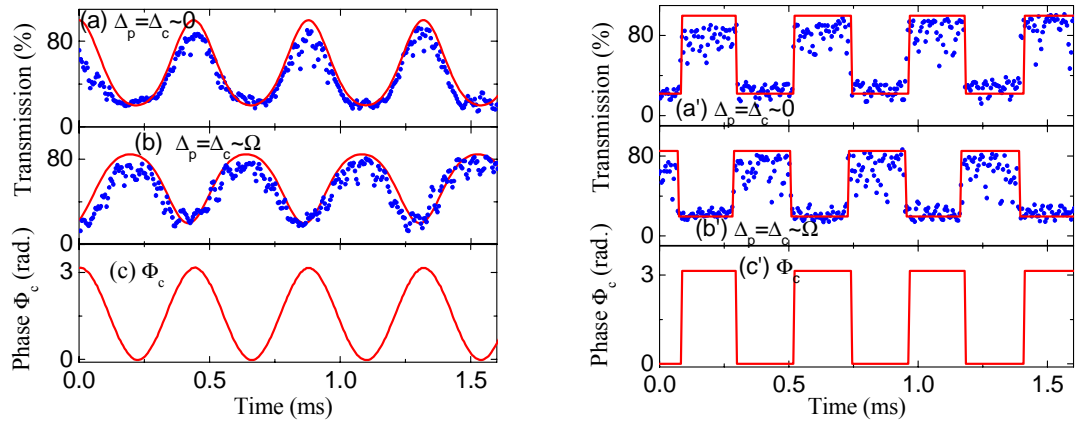

Fig. 5   Kang et al